\documentclass[aps,shownopacs,superscriptaddress,nofootinbib,preprint,11pt]{revtex4}
\usepackage{hyperref}
\usepackage{amsmath}
\usepackage{float}
\usepackage{graphics}
\usepackage{graphicx}
\usepackage{epsfig}
\usepackage{psfrag}
\usepackage{color}
\usepackage{slashed}

\usepackage{amsfonts}
\usepackage{amssymb}

\newcommand*{\be}{\begin{equation}}
\newcommand*{\ee}{\end{equation}}
\newcommand*{\bea}{\begin{eqnarray}}
\newcommand*{\eea}{\end{eqnarray}}

\newcommand{\comment}[1]{}
\newcommand{\esd}{m_{\tilde D_2}^2-m_{\tilde D_1}^2}
\newcommand{\est}{m_{\tilde t_2}^2-m_{\tilde t_1}^2}

\newcommand{\cref}[1]{Chapter~\ref{c.#1}}

\newcommand{\lab}[1]{\label{eq:#1}}
\newcommand{\barr}{\begin{eqnarray}}
\newcommand{\earr}{\end{eqnarray}}
\def\nn{\nonumber \\}

\def\beq{\begin{equation}}
\def\eeq{\end{equation}}
\def\bea{\begin{eqnarray}}
\def\eea{\end{eqnarray}}
\def\ba{\begin{array}}
\def\ea{\end{array}}
\def\bi{\begin{itemize}}
\def\ei{\end{itemize}}
\def\be{\begin{enumerate}}
\def\ee{\end{enumerate}}
\def\bc{\begin{center}}
\def\ec{\end{center}}
\def\bt{\begin{table}}
\def\et{\end{table}}
\def\btb{\begin{tabular}}
\def\etb{\end{tabular}}

\def\lsim{\raise0.3ex\hbox{$\;<$\kern-0.75em\raise-1.1ex\hbox{$\sim\;$}}}
\def\gsim{\raise0.3ex\hbox{$\;>$\kern-0.75em\raise-1.1ex\hbox{$\sim\;$}}}

\begin{document}

\title{A little more Gauge Mediation and the light Higgs mass}
\author{V. Suryanarayana Mummidi}
\email{soori9@cts.iisc.ernet.in}
\author{ Sudhir K  Vempati}
\email{vempati@cts.iisc.ernet.in}
\affiliation{Centre for High Energy Physics, Indian Institute of Science,
Bangalore 560012}

\begin{abstract}

We consider minimal models of gauge mediated supersymmetry breaking with an extra $U(1)$ factor in addition to the
Standard Model gauge group. A  $U(1)$
charged, Standard Model singlet is assumed to be present which allows for an additional NMSSM like 
coupling, $\lambda H_u H_d S$. The U(1) is assumed to be flavour universal. 
Anomaly cancellation in the MSSM sector requires additional coloured degrees
of freedom. The $S$ field can get a large vacuum expectation value along with consistent electroweak
symmetry breaking. It is shown that the lightest CP even Higgs boson can attain mass of the order
of 125 GeV. 
 \end{abstract}
\vskip .5 true cm

\maketitle
\section{Introduction}

Gauge mediated supersymmetry breaking \cite{Dine:1993yw,Dine:1994vc, Dine:1995ag,Giudice:1998bp} (for earlier works, 
please see, \cite{Dine:1981za,Dine:1981gu,Dine:1982zb,Dimopoulos:1982gm,Dimopoulos:1981au,Nappi:1982hm,AlvarezGaume:1981wy})
 attractive due to  several interesting features (i) flavour blind supersymmetry breaking soft terms (ii) very 
few parameters determine 
the entire spectrum (iii) different collider phenomenology compared to gravity mediated models as typically 
gravitino is the lightest supersymmetric particle (LSP) etc. However, phenomenologically\footnote{For an
early phenomenology of these models, please see, \cite{Agashe:1997kn,Bagger:1996bt,Baer:1996hx}. } 
the minimal versions of gauge mediation are severely constrained due to the discovery of a Higgs particle 
with a mass around 125 GeV.  In MSSM,  for the lightest CP even Higgs to be around 125 GeV would require, stop mixing parameter
 $X_t $ to be large, $X_t \sim \sqrt{6} M_S$, where $M_S = \sqrt{m_{\tilde{t}_1} m_{ \tilde{t}_2}}$. While this
 holds true as long as stops are light $\sim 1~ \text{TeV}$, for very heavy stops $\gtrsim 4\, \text{TeV}$, the
 mixing parameter can be smaller. This would however push stops out of the reach of the LHC. 
In spite of theoretically appealing features, unfortunately,  in minimal gauge mediation, the only way to fit a 
light Higgs mass $\sim 125 ~\text{GeV}$ is by making
 stops very heavy. The typical trilinear couplings in these models are very small at the mediation scale $\sim 0$. 
 Renormalisation group (RG) effects do generate them at the weak scale, however their magnitude is not 
 large enough unless one makes gluinos ultra heavy $\sim$ several TeV \cite{Draper:2011aa}. It should be noted that
 the constraints from 125 GeV Higgs boson are stronger even if one moves away from minimal mediation models
 to general gauge mediation models as long as $A_t$ remains zero at the messenger scale \cite{Grajek:2013ola}.

Several possible solutions have been explored in the literature \cite{Albaid:2012qk,Frank:2013yta,Evans:2013kxa,Craig:2013wga,Calibbi:2013mka, Byakti:2013ti, Evans:2011bea,Evans:2012hg,Craig:2012xp, Yanagida:2012ef, Jelinski:2011xe,Abdullah:2012tq, Perez:2012mj,Endo:2012rd,Martin:2012dg,Fischler:2013tva,Bhattacharyya:2013xma,Kang:2012ra}.  One of the directions which is popular with many authors
is to introduce direct Yukawa couplings between messenger fields and the MSSM fields in addition to gauge interactions \cite{Chacko:2001km,
Chacko:2002et}. In some cases, these interactions could also violate flavour \cite{Shadmi:2011hs}.  In most of the models it is possible 
to generate large enough $A_t$ at the weak scale to fit the 125 GeV light CP even Higgs boson mass. In a recent survey
 \cite{Evans:2013kxa,davidtalk} it has been pointed out that a particular class of messenger-matter interactions, messenger- stop 
 mixings,  has the least fine tuning of all the possible models which fit the light Higgs mass.
Another direction which has been considered is to add additional vector like quarks close to the weak scale  which couple to the
 Higgs superfields. These lead to additional corrections to the light Higgs boson thus lifting its mass without the need of increasing
 the stop masses (see for example,  \cite{Martin:2012dg,Endo:2012rd,Fischler:2013tva}). 

In the following we would like to take an alternate route. We would like to keep the minimal mediation structure
in tact, thus would not like to introduce direct couplings between matter and messenger fields.   Adding an additional 
 singlet field,  like in NMSSM could help to raise the light Higgs mass.  There are however, problems with electroweak
 symmetry breaking while  incorporating NMSSM  in minimal gauge mediation. These  are well documented in literature
  \cite{deGouvea:1997cx,Dine:1996xk}.  There are ways out, either by adding additional matter fields or dynamics through
  which NMSSM can be made viable with minimal gauge mediation \cite{ Langacker:1999hs,Ellwanger:2008py, Liu:2008pa,
  Hamaguchi:2011nm,Hamaguchi:2011kt,Dimopoulos:1997je,Morrissey:2008gm, Yanagida:1997yf,deBlas:2011cr,deBlas:2011hs}. 
Post 125 GeV Higgs boson, a  model within this class has been explored in \cite{Yanagida:2012ef}. 
  
  In the present work, we will consider an additional U(1) gauge group under which the `singlet' of the NMSSM is charged. 
  This U(1) factor also participates in gauge mediation. Anomaly cancellation requires additional vector like matter 
  to be present.  Such vector like matter is typically introduced to generate correct electroweak symmetry
  breaking while incorporating NMSSM in minimal mediation models \cite{Dine:1996xk}. In the present case, it is 
  motivated from anomaly cancellation requirements. It should be noted that this kind of model has been
  considered earlier by the authors of Ref. \cite{Langacker:1999hs}.  Ours is a more explicit realisation of it
  in the sense that we have taken care of $U(1)$ charges and anomaly cancellation conditions. Furthermore,
  we have performed a more detailed analysis of the Higgs masses in the light of 125 GeV Higgs discovery. 
      
 We found that it is possible to find an appropriate set of rational U(1) charges which satisfy the anomaly
 cancellation conditions as well as allow the correct set of terms in the superpotential.  Electroweak
 symmetry breaking is possible as the U(1) charged singlet can achieve a reasonable vacuum 
 expectation value (vev).   Two factors contribute to the raise in the lightest CP even Higgs mass: the 
 effective $\mu$ term is sufficiently large $\sim 0.5-1\, \text{TeV}$ and secondly the RG generated 
 $A_t$ term is large compared to minimal gauge mediation. The later is because at the 1-loop level, 
 the $SU(3)$ beta function,  $b_3$ is  zero in this model and the 2-loop $b_3$ is not sufficiently large. 
Together they result in sufficient $X_t $ to ensure large mixing in the stop mass matrix.  It is possible
to find reasonable parameter space which gives correct lightest CP even Higgs mass and satisfy 
direct constraints from LHC as well as constraints from $Z-Z'$ mixing.

The rest of the paper is organised as follows: In the next section particle spectrum and the model are presented. 
The details of supersymmetric spectrum and various constraints on the parameter space  are discussed in section 3. 
Numerical results are presented in section 4. We close with an outlook in section 5. 

\section{Model and the Particle Spectrum}

The basic premise of the model is that the singlet of the NMSSM should no longer be a singlet, but instead, it is  charged under
an extension  of the Standard Model gauge group such that it receives  non-zero supersymmetry breaking contributions at the mediation 
scale. As it will be detailed in the next section this would help in attaining a large enough vacuum expectation value for the field `S'. In this present work, 
we try to do this by considering the simplest extension in terms of a $U(1)$.  The relevant field
$S$ is singlet under the Standard Model gauge group, but charged under the extra $U(1)$ ; as  a consequence of which all the
Standard Model fields are charged under the $U(1)$. The total gauge group is 
\beq
G_{SM+A}  = SU(3)_c \times SU(2)_{L} \times U(1)_Y \times U(1)_A
\label{gsm}
\eeq
where the first three represent the usual Standard Model gauge group and the additional gauge group is represented by a subscript A. $U(1)_A$ is a chiral gauge group and hence
 introduces an extra set of anomalies which need to be canceled for a consistent quantum field theory. This imposes a set 
of conditions on the $U(1)_A$ charges; they are listed  in Appendix \ref{anomaly}.  We insist that the anomalies cancel independently for the NMSSM sector and the Messenger sector.
It is easily verified that the MSSM particle spectrum along with the new field $S$ is
not sufficient to cancel all the anomalies. In particular, from $(U(1)_A-[SU(3)_c]^2)$ anomaly condition we get
\beq
A_1(exotics)=-3s
\label{exotic}
\eeq
where $A_1 (exotics)$ is the contribution of the new exotic fields which need to be added and $s$ is the $U(1)_A$ charge of the field $S$. 
The $U(1)_A$ charge $s$  cannot be zero as per our requirements. Furthermore,  to generate the effective   $\mu$ term ($\lambda S H_u H_d $) in the super-potential, 
the charge $s$ should be equal to 
\beq
s=-(h_1+h_2) \neq 0
\eeq
where $h_1$ and $h_2$ are the $U(1)_A$ charges of $H_1$ and $H_2$ respectively. We thus need coloured exotics to satisfy $U(1)_A-[SU(3)_c]^2$ anomaly.
The number of the exotics is fixed by other anomaly conditions as well as by  the $U(1)_A$  gauge invariance of the  super-potential.  
It turns out that one possible minimal set of exotic fields would be three families of  $SU(2)_L$ singlet coloured exotics.  We introduce a pair of 
 colour fundamental and   anti-fundamentals $D_i$ and $\bar{D}_i$, which are $SU(2)$ singlets, for each of the three families. 
 In addition to the QCD interactions they are  allowed to couple with the field $S$  in the super-potential.  The total particle spectrum 
 and their corresponding representations and the $U(1)_A$ charges , in the order of Eq. (\ref{gsm})  are given in the table below.

\beq
\lab{qmnumbers}
\begin{array}{rlrlrl}
Q_i:& (3,2,\frac{1}{6},q_i)  & 
~U^c_i :& ({\bar3},1,-\frac{2}{3},u_i),\qquad
& D^c_i :& ({\bar3},1,\frac{1}{3},d_i), \qquad  \\
L_i:& (1,2,-\frac{1}{2},l_i),\qquad&~E^c_i: & (1,1,1,e_i),\qquad &   & \\ 
H_1:& (1,2,-\frac{1}{2},h_1) &
~H_2:& (1,2 ,\frac{1}{2},h_2), \qquad
&S:&  (1,1,0,s), \qquad \\
D_i:&(3,1,y_i,z_i) \qquad & \bar{D}_i&(\bar{3},1,-y_i,\bar{z}_i), \qquad &&
\end{array}
\eeq
where $i$ represents the generation index running from 1 to 3.  In the rest of the paper, we will consider all the $U(1)_A$ charges to be universal 
over all the generations and thus suppress the generation index. The only exception to this rule  is
 the $U(1)_A$ charges  of exotics $z_i$.  We will consider them to be different for each of the generation,  subject to the constraint 
 that in each generation,  $z_i + \bar{z}_i = -s $. The super-potential  is given by
\beq
W= Y_{E}L{E^c}H_1 + Y_{D}Q{D^c}H_1 +
Y_{U}Q{U^c}H_2+ \lambda SH_1H_2+ \kappa_i  S D_i  {\bar D}_i
\label{superpotential} 
\eeq
where $Y_E,\,Y_D,\,Y_U,\lambda$, $\kappa_i$ are Yukawa couplings and we have suppressed generation and colour indices.  Note that 
the field S does not have cubic self interactions. 

We will consider a minimal set of messengers communicating the effect of spontaneous supersymmetry breaking in the hidden sector.  The spurion $X$ couples to the
messengers with the super-potential
\beq
W = \eta  X \Phi\bar{\Phi}
\eeq
where  $\Phi$ are messengers in fundamental representation of an $SU(N) \supset G_{SM + A}$ gauge group and $\eta$ is some Yukawa
coupling.  The resultant soft terms can 
easily be generalised with the extra $U(1)_A$  and can be verified with the wave-function methods of Refs. \cite{Giudice:1997ni,ArkaniHamed:1998kj}. The mass terms for the gauginos and soft mass squared terms for the scalars at the mediation scale, $X$ are given as follows\footnote{In writing the formulae Eq.(\ref{bcs}) we have
suppressed the 1-loop and 2-loop functions. They are however taken in to account in the numerical analysis}:
\bea
M_i (X)&\approx&  \frac{\Lambda}{16\pi^2} \sum_i \left(g_i^2(X) \right) \nonumber  \\
m^2_{\tilde f} (X) &\approx& \frac{2\Lambda^2}{(16\pi^2)^2}\sum_i \left(g_i^4 (X) ~C_i (f) \right)
\label{bcs}
\eea
where through an abuse of notation, we have expanded the spurion as  $ <X> = X + \theta^2 F_X$ and 
defined $\Lambda = {F_X / X  }$. $ C_i(f)$ are quadratic Casimirs for the fields $f$ under the four gauge groups. The index $i$ here
runs over all the four gauge groups of Eq.(\ref{gsm}).  We denote the gauge coupling corresponding to $U(1)_A$ as $g_4$ and we
can see, the  soft mass of  $S$ has the following non-zero value   at the $X$ scale : 
\beq
m_{S}^2 (X) \approx    ~2 s^2 ~\tilde{\alpha}_4^2(X) ~ \Lambda^2,
\eeq
where we used the standard notation of $\tilde{\alpha}_i = \alpha_i /(4 \pi)$ and $\alpha_i = g_i^2 / (4 \pi)$. 
Similarly, we christen $M_4$  to be the neutral gaugino corresponding to $U(1)_A$ group.  It's mass is
given by 
\beq
M_4 \approx  \tilde{\alpha}_4(X)~ \Lambda
\eeq
The presence of additional $U(1)_A$ also introduces additional splittings between the mass squared terms at the mediation scale $X$. 
For example, the slepton doublets and the Higgs which  are degenerate at the high scale in Minimal case, get split as:
\bea
m_{L}^2 (X) - m_{H_{1,2}}^2 (X) & =& 2 (l^2 - h_{1,2}^2)~ \tilde{\alpha}_4^2(X) ~ \Lambda^2 \nonumber \\
m_{H_1}^2 (X) - m_{H_2}^2 (X) & = & 2 (h_1^2 - h_2^2) ~\tilde{\alpha}_4^2(X) ~ \Lambda^2
\label{splits}
 \eea
However, as we will see later the freedom of these splits is limited as the choice of  $U(1)_A$ is quite restricted due to phenomenological
constraints and anomaly cancellation conditions. 
Finally, just as in the minimal messenger model, the trilinear $A$ -terms and
bilinear $B$ terms  remain zero at the mediation scale $X$.  

\section{Weak Scale Spectrum}
The soft terms at the weak scale  can be evaluated by using the relevant Renormalisation Group (RG)  equations with the above boundary
conditions, Eq.(\ref{bcs}).
One interesting aspect about the one loop beta functions for the gauge couplings is that the beta function of $SU(3)$, $b_3^{(1)} =0$. 
This is due to  the presence of the additional colour
triplets $D, \bar{D}$ in three generations\footnote{ We have not explored in the present work about the possibility of making this model 
finite in the UV (see for example \cite{Babu:2002ki}).}.
As the $\alpha_s$ does not run at the 1-loop level, most coloured particles receive larger corrections in RGE running, compared to the 
Minimal messenger model. This has consequences for the running of $y_t$ and subsequently to all the parameters which 
depend on $y_t$ or $A_t$.
We have used 1-loop RGE for the soft terms and added  2-loop RGE's for the gauge couplings and Yukawa couplings in this analysis. The relevant RGE for this model are given in Appendix \ref{rge}. 

Before proceeding further, a comment about kinetic mixing is in order. The U(1) gauge fields can mix through the kinetic terms of the type 
$\chi \int d \theta~ \mathcal{W}^A \mathcal{W}_Y$. The current bounds on $\chi$ limit it to $10^{-3}$\cite{Hook:2010tw}. 
We  expect that the implications on the phenomenology to be discussed in our paper will be minimally affected due to the presence of the kinetic mixing. For this reason, we will neglect all its effects in the subsequent discussion. 

At the weak scale, $M_{SUSY} \sim 1 \,\text{TeV}$, we impose electroweak symmetry breaking conditions along with the $U(1)_A$ breaking. 
The neutral Higgs scalar potential is given by
\beq  
V_0 = V_F + V_D + V_{\rm soft}
\eeq
where
\bea
V_F &=& |\lambda H_2\cdot H_1|^2 + |\lambda S|^2 \left(|H_1|^2+|H_2|^2 \right), \\
V_D &=& \frac{(g_1^2+g_2^2)}{8}\left( |H_1|^2-|H_2|^2 \right)^2+ \frac{g_{2}^2}{2} \left( |H_1|^2|H_2|^2-|H_2 \cdot H_1|^2 \right)\\
 &+& {{g_{4}}^2\over2}\left(h_1 |H_1|^2+h_2 |H_2|^2+s |S|^2\right)^2\\ \nn
V_{\rm soft}&=&m_{1}^{2}|H_1|^2 + m_{2}^{2}|H_2|^2+ m_s^{2}|S|^2 +\left( A_\lambda S H_2\cdot H_1 + h.c. \right).
\label{eq:potential}
\eea
The neutral components of the Higgs fields $H_1$ and $H_2$ get vacuum expectation values (VEV) at the weak scale, 
$\frac{ v_1} {\sqrt{2}}$ and 
$\frac{ v_2}{ \sqrt{2}}$. The  field  $S$ also gets a  VEV,   $\frac{v_s}{\sqrt{2}}$ at the weak scale, breaking the $U(1)_A$ symmetry spontaneously.  
At the minima of the potential, the \text{vev}s and  the soft terms along with the other parameters of the model get related. These minimisation conditions
are given as 
\begin{eqnarray}
\label{min1}
m_1^{2}&=& - \frac{1}{2}\left[\frac{G^2}{4} + h_1^{2} {g_{4}}^{2}\right] v_1^{2} + \frac{1}{2} \left[\frac{G^2}{4} - \lambda^{2} - h_1 h_2 {g_{4}}^{2}\right] v_2^{2} - \frac{1}{2} \left[\lambda^{2} + h_1 s {g_{4}}^{2}\right] {v_s}^{2} \nonumber \\
&& +\ \frac{A_\lambda}{\sqrt{2}} \frac{v_2 v_s}{v_1}, \\
\label{min2}
m_2^{2}&=& \frac{1}{2} \left[\frac{G^2}{4} - \lambda^2 - h_1 h_2 {g_{4}}^{2}\right] v_1^{2} - \frac{1}{2} \left[\frac{G^2}{4} + h_2^{2} {g_{4}}^{2}\right] v_2^{2} - \frac{1}{2} \left[\lambda^{2} + h_2 s {g_{4}}^{2}\right] {v_s}^{2} \nonumber \\
&& +\ \frac{A_\lambda}{\sqrt{2}} \frac{v_1 v_s}{v_2}, \\
\label{min3}
m_s^{2}&=& - \frac{1}{2} \left[\lambda^{2} + h_1 s {g_{4}}^{2}\right] v_1^{2} - \frac{1}{2} \left[\lambda^{2} + h_2 s {g_{4}}^{2}\right] v_2^{2} - \frac{1}{2} s^{2} {g_{4}}^{2}v_s^{2} + \frac{A_\lambda}{\sqrt{2}} \frac{v_1 v_2}{v_s},
\end{eqnarray}
 where $G^2 = g_1^2+g_2^2$. The minimisation conditions are modified compared to the standard NMSSM case due to the presence of terms proportional
 to $g_4$.  Subsequently,  we can see from Eq. (\ref{min3}), that in the limit $v_s \gg v_1, v_2$($v_s$ is required to be large which is discussed later in this section), we have  $$ v_s^2 \approx - {2 ~m_s^2 \over s^2 g_4^2}, $$ 
 which is the typical \textit{vev} one expects in extra U(1) models \cite{Barger:2008wn,Langacker:1999hs}. At the high scale, $X$, $m_{S}^2$ which is positive
  and proportional to $\tilde{\alpha}_4^2 \Lambda^2$ can  be driven negative at the electroweak scale by the  Yukawa couplings of the exotics $k_1,k_2,k_3$ .
  
  This should be contrasted with the \text{vev} in minimal gauge mediation, without the $U(1)$ factor. See for example,Refs.[ \cite{deGouvea:1997cx,Ellwanger:2009dp} ].  From the minimization conditions of NMSSM,  we get
 \begin{equation}
 v_s^2 \approx -\frac{1}{2\kappa^2}\left(\lambda^2 (v_1^2+v_2^2) +2m_s^2-2\lambda \kappa v_1 v_2\right)
 \end{equation}
 which is too small to get $\mu_{eff}$ ($\frac{\lambda v_s}{\sqrt{2}}$) of the order of electroweak symmetry breaking. To achieve a significant value
 either $\lambda$ has to be very large ($>1$) or $\kappa$ has to be too small. In both the cases, achieving 
 electroweak symmetry breaking is highly constrained \cite{Delgado:2007rz}. 
We now turn our attention to the Higgs sector.
The CP-even tree-level Higgs mass squared matrix, $\Psi^\dagger \mathcal{M}_{+}^2 \Psi $, where $\Psi^T = \{H_1^0, H_2^0, S\}$,
and the elements of the matrix are given as:
\begin{eqnarray}
\left({\mathcal{M}_{+}^0}\right)_{11}^2 &=&  \left[\frac{G^2}{4} + h_1^{2} {g_{4}}^{2}\right] v_1^2 +\frac{A_\lambda}{\sqrt{2}} \frac{v_2 v_s}{v_1}\nn
\left({\mathcal{M}_{+}^0}\right)_{12}^2 &=& -\left[\frac{G^2}{4} - \lambda^{2} - h_1 h_2 {g_{4}}^{2}\right] v_1 v_2 -\frac{A_\lambda}{\sqrt{2}}v_s\nn
\left({\mathcal{M}_{+}^0}\right)_{13}^2 &=&  \left[\lambda^{2} + h_1 s {g_{4}}^{2}\right] v_1 v_s - \frac{A_\lambda}{\sqrt{2}}v_2\nn
\left({\mathcal{M}_{+}^0}\right)_{22}^2 &=&  \left[\frac{G^2}{4} + h_2^{2} {g_{4}}^{2}\right] v_2^2 + \frac{A_\lambda}{\sqrt{2}}\frac{v_1 v_s}{v_2} \nn
\left({\mathcal{M}_{+}^0}\right)_{23}^2 &=&  \left[\lambda^{2} + h_2 s {g_{4}}^{2}\right] v_2 v_s - \frac{A_\lambda}{\sqrt{2}}v_1\nn
\left({\mathcal{M}_{+}^0}\right)_{33}^2 &=&  s^{2} {g_{4}}^{2}v_s^2 +\frac{A_\lambda}{\sqrt{2}}\frac{v_1 v_2}{v_s}
\label{cpevenHiggs}
\end{eqnarray}

\begin{figure}[h]
 \includegraphics[width=8cm]{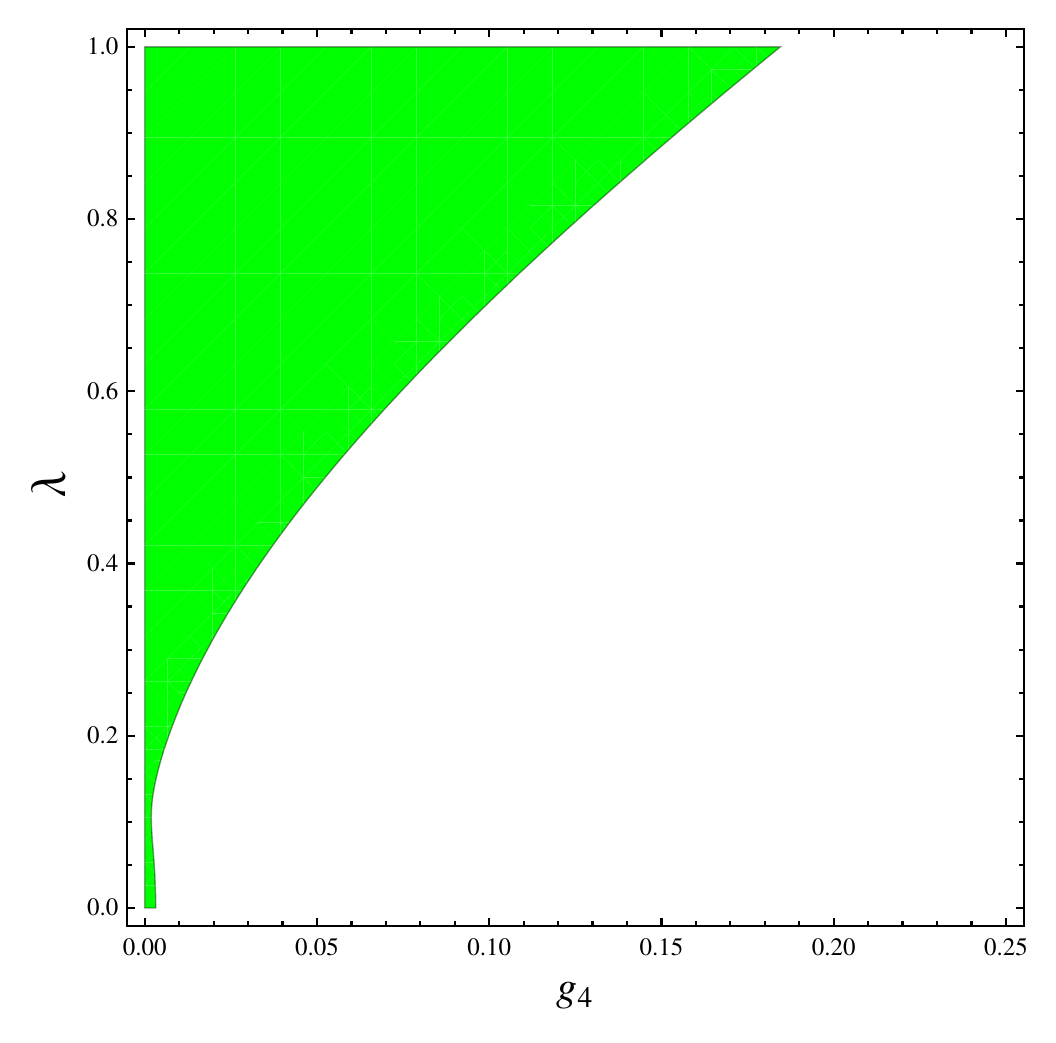}
  \caption{The determinant of the CP even Higgs mass matrix is shown as a function of $g_4$ and $\lambda$.  In the  shaded region, the
  determinant is negative, thus electroweak symmetry breaking is not possible. The $U(1)$ charges used are presented in Table \ref{charges} and
  tan$\beta$ is chosen to be 10.   }
  \label{det}
\end{figure}

Given that the physical Higgs spectrum should be non-tachyonic, we can derive constraints on the parameter space
of the model. 
Firstly the sign of the determinant of the matrix, in the limit $v_s >> v_{1,2}$ is crucially dependent on the 
sign of the $A_\lambda$. This is obvious, by considering the full determinant of the $3 \times 3$ mass matrix, 
which is given by 

\begin{eqnarray} 
Det[(\mathcal{M}_{+}^{0})^2] &\approx& \frac{A_\lambda v_s^3}{4 \sqrt{2} v_1 v_2}\left[G^2\, g_4^2\, s^2\, (v_1^2 - v_2^2)^2 + 4 \,\left(g_4^4 \,h_1^2\, s^2\, v_1^4 - (\,l^4 + 2\, g_4^2\, l^2 \,(\,h_2 - s\,)\, s + g_4^4\, h_2\right .\right.\nn
&&\left.\left. (-2\, h_1 + h_2\,)\, s^2)\, v_1^2 v_2^2 + g_4^4 h_2^2 s^2 v_2^4\right)\right]\nonumber
\end{eqnarray}
For $A_\lambda >0$, the region in which the sign of the determinant of the Higgs mass matrix changes is plotted in $\lambda,g_4 $ plane by taking  $ h_1=-\frac{1}{2},h_2=-\frac{5}{2},s=3$, and $\tan{\beta}=10 $. Electroweak symmetry breaking is not possible for the shaded region ($Det< 0$) in the parameter space. From the figure, it is seen that for $g_4 \lesssim 0.1 $,
large values of $\lambda \gtrsim 0.6$ are disfavoured as they do not allow electroweak symmetry breaking. 

\begin{figure}[h]
\includegraphics[width=0.45\textwidth,angle=0]{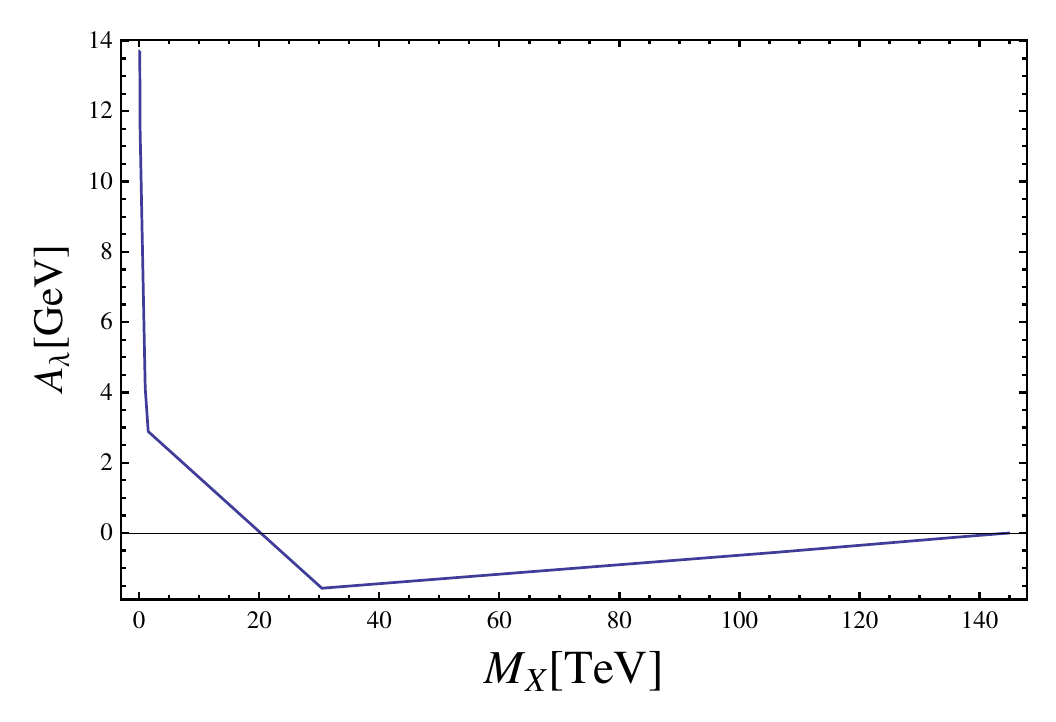}
\includegraphics[width=0.45\textwidth,angle=0]{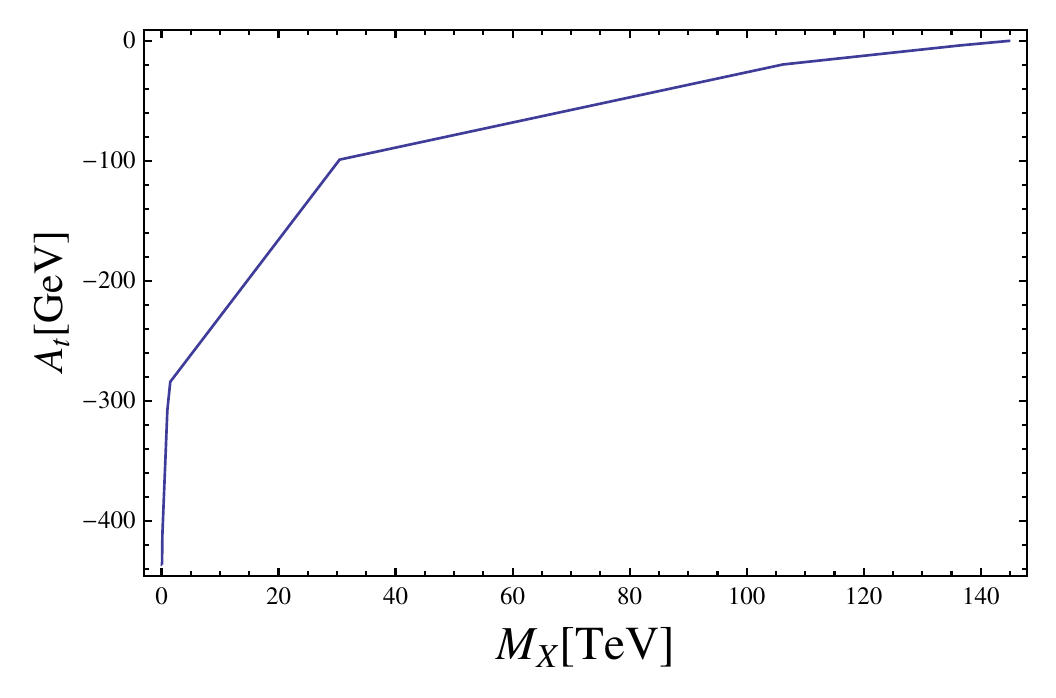}
\caption{$A_{\lambda}$ and $A_t$ are plotted as a function of the energy scale, where free parameters are fixed as $\lambda=0.394$,~$g_4=0.137$,~$k_1=0.016$,~$k_2=1.07$,~$k_3=0.117$,$\tan{\beta}=3.7$ }
\label{alambdarunning}
\end{figure}

The question then arises, whether $A_\lambda >0 $ ?. 
Typically the A terms are negative due to the RG running from the high scale. However, in this case,
$A_\lambda$ turns out to be $\mathcal{O}(10)$ and positive at the weak scale. 
This positive $A_\lambda$ ensures us a safe electroweak vacuum. This is shown in the left panel of Figure $\ref{alambdarunning}$ , where
we have plotted $A_\lambda$ with respect to running scale.
As we see from the figure  $\ref{alambdarunning}$, $A_{\lambda}$ initially turns negative and then increases turning positive at the 
weak scale. This happens because of the complicated coupling between $A_t$ and $A_\lambda$ RGE. 
The RGE of these parameters are presented in the Appendix \ref{rge} along with the other parameters. In the
below, we reproduce them:
\begin{eqnarray}
{d A_t \over dt} &\approx&\frac{y_t}{16 \pi^2}\left[2y_bA_b+A_{\lambda}\lambda+{32\over3}g_3^2M_3+6g_2^2M_2+{26\over9}g_1^2M_1+4(q^2+u^2+h_2^2)g_4^2M_4\right] \nonumber \\
\cr\frac{d A_\lambda}{dt} &\approx& \frac{\lambda}{16 \pi^2}\left[6y_tA_t+6y_bA_b+2A_\tau y_\tau+6(A_{k_1}k_1+A_{k_2}k_2+A_{k_3}k_3)\right.\nn
&&\left.+6g_2^2M_2+2g_1^2M_1+4(s^2+h_2^2+h_1^2)g_4^2M_4\right]\nonumber
\end{eqnarray}
Compared to the minimal gauge mediated models, the running effects on the parameter $A_t$ are very large  
as $\alpha_3$ barely runs in this models.  As mentioned above, $b_3 = 0$ at 1-loop and is very small, at the 2-loop. For this reason, after the SUSY threshold $M_S \sim 1 \text{TeV}$,  $\alpha_s$ barely runs 
 all the way to the mediation scale.
Due to this $Y_t$ and $A_t$ receive comparatively large corrections due to the relatively  large $\alpha_s$. 
Additional corrections from $g_4, k_i$ and $A_{k_i}$ also contribute in the running of the $A_\lambda$.
This feeds into $A_\lambda$, making it positive at the weak scale.  In the right panel of the Fig [\ref{alambdarunning}], we show the running of the $A_t$ for the same parameters  

Let us focus our attention to the lightest Higgs mass eigenvalue.  The matrix Eq.(\ref{cpevenHiggs}) gives
an upper bound on the tree level  lightest Higgs mass. In the present model, it  has additional 
contribution from $\lambda$ and $g_4$ which is given as
\begin{equation}
{m_{h_0}}^2 \le M_Z^2 \left[\cos{2\beta}^2+\frac{\lambda^2}{2 g^2}\sin{2\beta}^2+\frac{g_4^2}{g^2}(h_1+h_2+(h_1-h_2)\cos{2\beta})^2\right]
\end{equation}
In the NMSSM, it is well known that the tree level contribution can be appreciably  enhanced from the MSSM tree level values
only for large values of $\lambda \gtrsim 0.7$.  The above bound is thus saturated only for special values
of the parameters. For most of the parameter space, however the actual eigenvalue is far below the above
bound. As in MSSM, one loop corrections would play a major role.

The total number of parameters are $\Lambda$, $M_X$, $g_4$, $\lambda $ and the $U(1)$ charges.  Before proceeding
to present the numerical results, we discuss the possible constraints  on the various parameters.  The first
constraint we discuss is from the neutral gauge boson mixing. 
The neutral gauge bosons $Z$ and $Z'$  mix with their mass matrix given by
  $\mathcal{L} ~\supset ~\chi^T \mathcal{M}_{Z'Z}^2 \chi $ 
where $\chi^T = \{ Z',Z \}$
with 
\begin{equation}
\mathcal{M}_{Z'Z}^2 = \left( \begin{array}{cc} M^2_{Z'Z'} & M^2_{Z'Z} \\
M^2_{Z'Z} & M_{ZZ}^2  \end{array} \right) 
\end{equation}
where 
\begin{eqnarray}
M^2_{Z'Z'} &=& {g_{4}}^2(h_1^2 v_1^2+h_2^2 v_2^2+s^2 v_s^2), \nonumber \\
 M^2_{ZZ'} &=& {g_{4}} \sqrt{g_1^2+g_2^2} \left( v_1^2 h_1 - v_2^2 h_2\right), \nonumber \\
 M^2_{ZZ}& =& {({g_1^2+g_2^2})\left( v_1^2+v_2^2\right)\over4}.
\end{eqnarray}
The mixing of the matrix is given by 
\begin{equation}
\Theta_{ZZ'} = {1\over2} \tan^{-1} \left( {2 M^2_{ZZ'} \over M_{Z'}^2 - M_Z^2}\right).
\end{equation}
The current limits on $M_{Z'}$ require it to be greater than 1 TeV \cite{Chatrchyan:2012ku}.  For $g_4 \sim g_1$, these limits already push 
 $v_s$ to be much larger than 1 TeV. $\Theta_{ZZ'}$ is constrained by electroweak precision data, it should be less than
  $O(10^{-3})$ \cite{Hook:2010tw}.  As $v_s$ is already very heavy with  $M_Z'$ of a  mass of TeV order,   the constraint 
  on mixing angle is avoided easily.

A second constraint comes from the mass spectrum of the scalar super-partners. 
The D-terms due to the new $U(1)_A$ group play an important role in determining the  sfermion mass spectrum 
due to the large vev of the $S$ field. The strongest effects are felt in the stau mass matrix which is given as:  
\bea
\mathcal{M}_{\widetilde{\tau}}^2  = \left(\begin{array}{cc} m_{\widetilde{L_3}}^{2}+m_{\tau}^2+ D_L  &  {1 \over{\sqrt{2}}} 
(A_{\tau}v_1-{\mu}y_{\tau} v_2 )\\
 {1 \over{\sqrt{2}}} ( A_{\tau}v_1-{\mu}y_{\tau} v_2)
 & m_{\widetilde{e_3}}^{2}+m_{\tau}^2+D_e\end{array}\right),
\eea
where 
\bea
D_L&=&{1 \over 8}(v_1^2-v_2^2)(-g_2^2+g_1^2) + {1 \over 2} g_4^2 l~(h_1 v_1^2+h_2 v_2^2+s v_s^2)\\
D_e&=& -{ 1 \over 4} (v_1^2-v_2^2)~g_1^2 +{1  \over 2} g_4^2   e(h_1 v_1^2+h_2 v_2^2+s v_s^2). 
\eea
Notice that for the $D_L$ and $ D_e$ to have positive values, the products of the $U(1)_A$ charges, $ls$ and $es$ 
should always be positive.  This is because unlike $m_{\widetilde{Q}}^2,m_{\widetilde{u}}^2$ and $m_{\widetilde{u}}^2$, 
the value of $m_{\widetilde{L}}^2$ at electroweak scale due 
to running is very low, as it should be, owing to the fact that $y_\tau \ll y_t$. So the sign of the diagonal terms in the stau 
mass squared  matrix depends on the $D_L$ and $D_e$ which in turn depends on the dominant term  $l\,s \,g_4^2 \,v_s^2$.
If we choose $ U(1)_A $ charges $l$ and $s$ of different signs we expect tachyonic masses for stau's. 
 
 The chargino mass matrix remains unaltered compared to the MSSM whereas the neutralino mass matrix is now expanded to include the neutral
 gauging of $U(1)_A$ as well as the fermionic partner of the $S$ field. Note that the fermonic partner of the $S$ is not 
 exactly the singlino as it carries a $U(1)_A$ charge unlike the NMSSM case.  To summarise the constraints, we have :
 
 \begin{itemize}
 \item  \textit{For consistent electroweak breaking :} we need,  $\lambda$ $ =\sqrt{2}\,\, \frac{\mu_{eff}}{v_s} $ and $A_\lambda >0$. So $\lambda$ cannot be arbitrarily large for a given $g_4$ which is evident from the Figure $\ref{det}$
 \item \textit{From $Z-Z'$ mixing:} we require that $v_s \sim $ \,$O(TeV)\gg v_1,v_2$
 \item \textit{Sfermion masses}: From the $D$-terms of the sfermion mass matrices, we require that $U(1)_A$ charges $l$ and $s$ should have opposite signs 
 \item Landau pole:the new gauge coupling 
\begin{equation}
 g_4 < 2\pi \sqrt \frac{2}{b_4 \log\frac{M_X}{M_z}}.\nn
\end{equation}
 $g_4 \approx 0.28$ for $b_4=145$ and\,$M_X=100$ \,TeV
 \end{itemize}
 
\section{Numerical Results}

\begin {table}%
\caption {$U(1)_A$ charges of the fields}
\begin {tabular}{|c| c| c|c | c|c|c| c|c| c| c|c | c| c |c |}
\toprule %
q& u& d &l &e &$h_1$&$h_2$ & s& $z_1$ &$z_2$&$z_3$& $\bar{z_1}$ &$\bar{z_2}$&$\bar{z_3}$ \\ \hline
${1\over 6}$& ${1\over 3}$& ${7\over 3}$&${1\over 2}$&${2}$&$-{5\over2}$&$-{1\over 2}$&${3}$&-3&${-1}$&${-1}$&${0}$&-2&-2 \\ [1ex]\hline
\end {tabular}
\label{charges}
\end {table}
To compute the  sparticle spectrum at the weak scale, we use a modified version of the publicly available code
SuSeFLAV\,\cite{Chowdhury:2011zr} with 2-loop RGE for the gauge couplings and the Yukawa couplings. The RGE for the
rest of the soft parameters are evaluated at the 1-loop level. For the
Higgs spectrum, we compute the full 1-loop effective potential corrections presented in Appendix D.  These corrections come from stop-top loop and the exotic quarks loop. Stop-top loop correction is the dominant contributor to the Higgs mass at the one loop. The correction due the exotic quarks is significant. It changes Higgs mass  by few percent and we have checked that it is possible to get Higgs mass of 125 GeV by adding both the corrections, although we have not considered exotic quarks loop correction to the Higs mass in the numerical analysis. 
The free parameters are $\Lambda$,~$\tan{\beta}$,~$\lambda$,~$g_4$,~$k_1$,~$k_2$ and $k_3$. These are randomly fixed at the low energy scale, for each set of these parameters, using RGEs we obtain corresponding values at the GMSB scale $X\simeq\Lambda$. Now along with the boundary conditions for the soft masses and A-terms, the same parameters are run down to the electroweak scale to check whether they satisfy minimization conditions given in section (2) and other constraints presented in section (3). This process is repeated several times to obtain a parameter space which satisfy electroweak symmetry breaking conditions. Subsequent to this, we impose phenomenological constraints 
from  direct SUSY searches  at LHC \cite{atlas:12,cms:12} 
as well as the flavour constraints from $b \to s + \gamma$ and $b \to s+ \mu^+ \mu^-$.  

In the numerical analysis, we fix the $U(1)_A$ charges to be as given in Table \ref{charges}. It should be noted that
these are not the only solutions available from anomaly cancellation conditions.  A list of five solutions is presented in 
 Appendix \ref{anomaly}. Of the remaining
parameters, we have fixed $\tan\beta = 10 $ and varied the remaining parameters within a range presented 
in  Table (\ref{parameter ranges}). 

\begin {table}[h]
\label{parameter ranges}
\caption {Ranges for the various Parameters}
\begin {tabular}{|c| c| }
\toprule %
Parameter & Range \\ \hline \hline
$\Lambda$ & $1\times10^5\,-\,5\times 10^7 [GeV]$\\ \hline 
$g_4$ & $0.01-2.5$ \\ \hline 
$\lambda$ &$ 0.1-0.9$ \\ \hline
$ \kappa_1$ & $ 0.1-0.9$\\ \hline 
$ \kappa_2 $& $ 0.1-0.9$\\ \hline 
$ \kappa_3 $&$ 0.1-0.9$ \\ \hline  
\end {tabular}
\end {table}

\begin{figure}
\includegraphics[width=0.45\textwidth,angle =0]{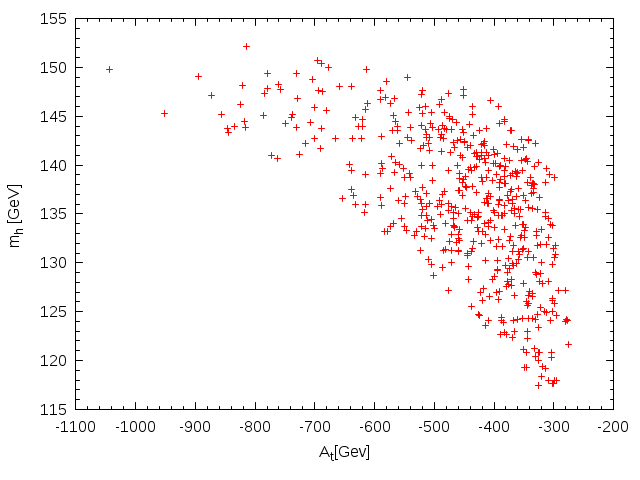}
 \includegraphics[width=0.45\textwidth,angle =0]{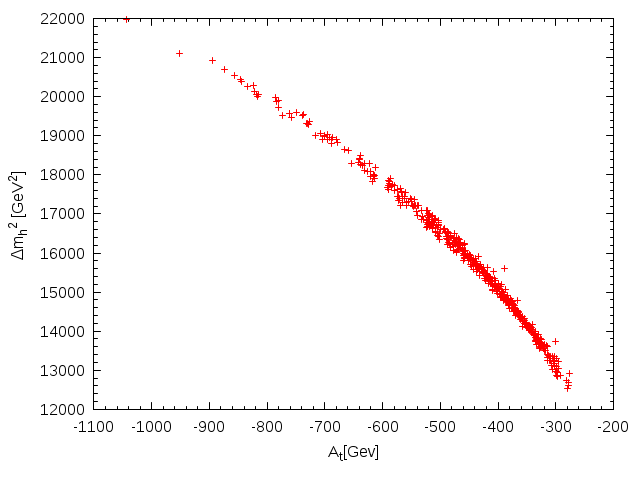}
 \caption{Higgs mass, including one-loop correction, and only one loop correction are plotted against $A_t$. The U(1) charges
 are taken from Table \ref{charges}. }
 \label{corr1}
 \end{figure}

Instead of presenting the results in terms of regions of allowed  parameter space,  we present the correlations of the
parameters with the lightest CP even Higgs boson mass. In Fig. (\ref{corr1}), we present the correlation of the 
light Higgs mass with respect to the $A_t$ generated at the weak scale.  The left panel presents the total Higgs mass
whereas the  right panel shows the 1-loop correction to the light Higgs mass. As expected we see that as $|A_t|$ increases,
the 1-loop correction to the Higgs mass increases so does the total mass. It is also surprising to see larger values for $A_t
\sim 900~\,GeV$ possible in this case and accordingly the higher values for Higgs mass $\sim 140 ~\,GeV$.  Of course, the heavier
Higgs masses correspond to heavier stops. Note that we have
considered only dominant 1-loop corrections to the light Higgs mass. Two loop contributions \cite{Martin:1993zk} 
 can be important and they would give a more precise number for the light Higgs mass.  However, it is clear that one 
 can easily achieve a light Higgs mass of  $\mathcal{O}(125)\, \text{GeV}$. 

\begin{figure}
 \includegraphics[width=0.45\textwidth,angle=0]{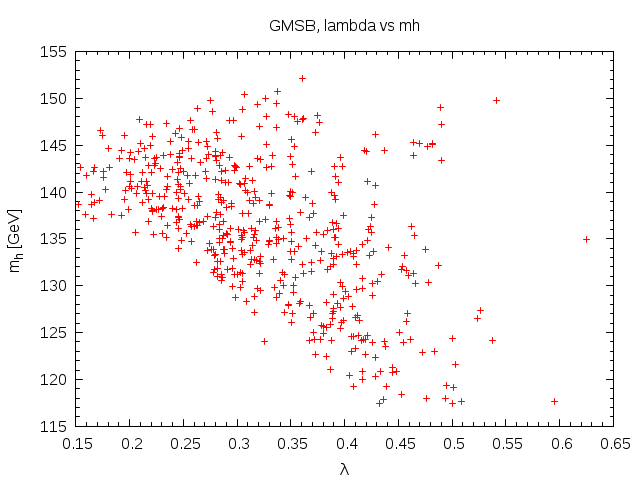}
\includegraphics[width=0.45\textwidth,angle=0]{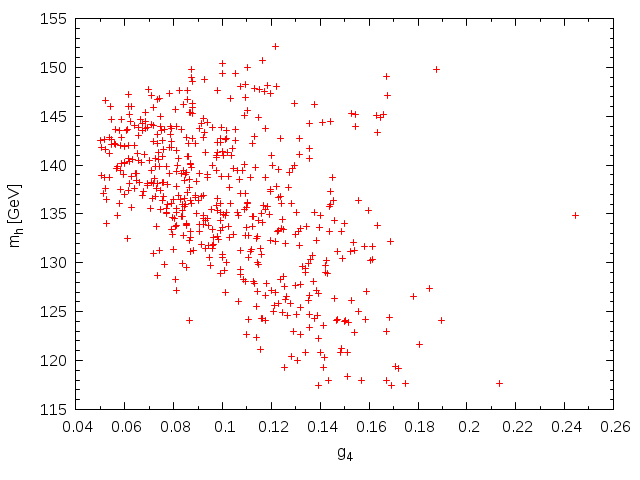}
\caption{Higgs mass, including one-loop correction is plotted against $\lambda$ and $g_4$ }
\label{corr2}
\end{figure}

 In Fig. (\ref{corr2}), we present the correlation between $m_h$ and $\lambda$ in the left panel and $m_h$ and $g_4$
 in the right panel. We find a surprising relation between $\lambda$ and $m_h$.  The Higgs mass seems to be lower for
 higher values of $\lambda$.  This is contrary to expectations based on NMSSM. This is because for higher values
 of $\lambda$ achieving electroweak symmetry breaking becomes harder.   Similarly, larger values of $\lambda$ typically
 mean lighter values of $v_s$. Similarly,  larger values of $g_4$ are not preferred by the data as they can lead
 to Landau poles.  This can be seen from  the right panel of Fig.(\ref{corr2}).  Thus, the regular NMSSM like enhancement
 of the tree level Higgs mass is not possible in this model. 
  
From the allowed parameter space, we now present a representative  point,  Point(A) which give the lightest
Higgs mass to be around 125 GeV.  In this point,  the next to  lightest supersymmetry particle (NLSP) is the A-ino, 
the  supersymmetric partner for the extra $U(1)_A$ gauge boson.

\noindent 
Point (A): \\  
\noindent 
The various  parameters for this point are :  $v_s=2225.53 \text{GeV} $,~$\tan(\beta)=3.26$,\,$\lambda=0.3439$,\,$g_4=0.1198$,~$M_X=194.22 ~\text{TeV}$,
$\Lambda =97.112  \text{TeV} $, ~$\kappa_1=0.1368$, ~$\kappa_2=0.7865$,~$\kappa_3=0.7813$
\begin{center}
\begin{tabular}{|c | c |c |c |c |c |}
\toprule
 Parameter & mass(GeV) & Parameter & mass(GeV) & Parameter & mass(GeV) \\ \hline
 $\widetilde{t_1}$&773.35   & ${\chi}_1^0$ & 37.00     & $ h_1^0$     &127.1     \\ \hline
 $\widetilde{t_2}$& 882.39   & ${\chi}_2^0$ & 122.26     & $ h_2^0$     &244     \\ \hline
 $\widetilde{b_1}$& 847.4  & ${\chi}_3^0$ & 544.8    & $ h_3^0$     &802.8     \\ \hline
 $\widetilde{b_2}$&1002.5   & ${\chi}_4^0$ & 554.19    & $ A^0$     &370.99     \\ \hline
 $\widetilde{{\tau}_1}$& 294.25& ${\chi}_5^0$ &799.7    & $ {\chi}_1^{\pm}$   &123.16     \\ \hline
  $\widetilde{{\tau}_2}$&460.58& ${\chi}_6^0$ &806.8    & $ {\chi}_2^{\pm}$  &549.94    \\ \hline
   $\tilde g$&911.5& $A_\lambda$ &10.1    & $ A_t$  &-279.3    \\ \hline
\end{tabular}
\end{center}

\section{Outlook}

The discovery of a Higgs boson at 125 GeV has led to strong constraints on the gauge mediated
supersymmetry breaking models.  Most of the present models have concentrated on generating
the required large trilinear  $A_t$ coupling through messenger matter interactions. In the present
work, we tried a different approach of  combining the ideas of an extra $U(1)$ factor and
NMSSM like models.  Anomaly cancellation requirement automatically determines the extra
particle spectrum of the model.  The coloured particles barely run in this model from the weak
scale to the mediation scale due to the small value of the strong beta function. This `stagnation'
of $\alpha_s$ between $M_{SUSY}$ and  $M_{mess}$ and the presence of additional U(1) 
couplings helps for a  larger value  of the $A_t$ at the $M_{SUSY}$ even though one starts with 
zero at the mediation scale. Together with a reasonable value for the $\mu_{eft} = \lambda v_s$, 
this generates the required $X_t$ at the weak scale for the light stops. 

While we have focussed on getting the right Higgs mass, the rest of the spectrum of the model
is also quite interesting. There are heavy  exotic coloured particles, new neutralinos which 
are combinations of the  Standard Model singlino and the fermion of the $U(1)_A$ gauge boson. 
The lightest neutralino is  still the LSP and could be the dark matter candidate.  A study of 
collider signatures and dark matter issues could be interesting and will be pursued in a future work. 

Finally, we have not concentrated on the issue of fine tuning in this model. Though we have not 
explicitly measured it,  it is expected that it could be large as long as $M_X$ and $\Lambda$ are close
as we have chosen.  A reasonable separation between the scales can perhaps reduce the
fine tuning (see for example, discussion in \cite{Komargodski:2008ax}).

\vskip 2 true cm 

\noindent 
\textbf{Acknowledgments}\\
We acknowledge discussions and important inputs from E. J. Chun. We also acknowledge discussions with
 P. Bandopadhyaya.  We also thank L. Calibbi to bringing to our notice a reference. SKV is supported by DST Ramanujan Grant SR/S2/2008/RJN-25 of Govt of India. VSM is supported by CSIR fellowship 09/079(2377)/2010-EMR-1.

\appendix
\section{Anomaly Conditions}
\label{anomaly}
In the following we present the anomaly cancellation conditions and U(1) charges which are solutions to them. 
More elaborate analysis of anomaly cancellations pertinent to U(1) extensions of MSSM has been presented in \cite{Lee:2007fw}. 
To begin with, the $U(1)_{A}$  gauge invariance of  the superpotential Eq.(\ref{superpotential})  leads to the below equations
which should be satisfied by the $U(1)_{A}$ charges. 
\bea
h_1+q+d &=& 0 \label{yd}\\
h_2+q+u &=& 0\label{yu}\\
h_1+l+e &=& 0 \label{yl}\\
s+h_1+h_2 &=& 0 \label{ys}
\eea
In addition, the following five anomaly cancellation conditions should also be satisfied. 
\begin{eqnarray}
\mathcal A_1 &:& U(1)_A-[SU(3)_C]^2\nn
\mathcal A_2 &:& U(1)_A-[SU(2)_L]^2\nn
\mathcal A_3 &:& U(1)_A-[U(1)_Y]^2 \nn
\mathcal A_4 &:& U(1)_Y-[U(1)_A]^2 \nn 
\mathcal A_5 &:& U(1)_A^3\nonumber
\end{eqnarray}
In the following, we analyse each of these conditions and the corresponding solutions for $U(1)_{A}$ charges. 
\subsubsection{Anomaly $\mathcal A_1 (U(1)_A-[SU(3)_C]^2)$}
\beq
3(2q+u+d)+\mathcal {A}_1(exotics)=0 \label{A1}
\eeq
Here  first term is the contribution from three generations of the quarks in the MSSM without considering the exotic 
$D,\bar{D}$ quarks presented in section (1).  We can show in the limit $\mathcal{A}_1(exotics) ~= 0$, the $S$ field
$U(1)_A$ would go to zero. This can be easily seen by  considering  the combination of the equations:
Eq.(\ref{A1}) - 3 Eq.(\ref{yu})  - 3 Eq.(\ref{yd}) + 3 Eq.(\ref{ys}), gives us
\beq
\mathcal  A_1(exotics)=-3s
\label{exotic}
\eeq

We assume that the exotics  are triplets and anti triplets of $SU(3)_c$ with equal and opposite $U(1)_Y $ hypercharges $\pm y_i$. 
Eq. (\ref{A1}) now becomes
\beq
3(2q+u+d)+\Sigma_i(z_i+{\bar z}_i)=0 \label{A11}
\eeq
where $z_i$ are the $U(1)_A$ charges of the exotics. The coupling between the exotic vector like quarks the singlet is allowed
under $U(1)_A$ symmetry which gives 
\beq
s+z_i+{\bar z}_i=0 \label{yk}
\eeq
Finally, to derive the number of families of exotic quarks one should add, consider the combination 
Eq. (\ref{A11}) - 3 Eq. (\ref{yu}) - 3 Eq. (\ref{yd}) + 3 Eq. (\ref{ys})- $\Sigma_i$ Eq. (\ref{ys}).  We have  $(3-N_k) s=0$, 
where $N_k$ is the number of exotic families which ends up being equal to three. 

\subsubsection{Anomaly $\mathcal A_2 (U(1)_A-[SU(2)_L]^2)$}
The constraint here is given as 
\beq
9\,q+3\,l+h_1+h_2 =0 \label{A2}
\eeq
From Eqs. ($\ref{yd}$), ($\ref{yu}$), ($\ref{yl}$), ($\ref{ys}$)  and  ($\ref{A2}$) we have 5 constraints.  Without the $U(1)_A$ charges
of the exotics, we have eight unknowns. Using the constraints,  a general solution can be written in terms of $l,h_1,s$ as
\bea
\label{gs}
\left(
\begin{array}{c}
 q \\
 u \\
 d \\
 e \\
 h_2\\
\end{array}
\right)={l\over3}
\left(
\begin{array}{c}
 -1 \\
 1  \\
 1  \\
-3 \\
 0 \\
\end{array}
\right)+h_1
\left(
\begin{array}{c}
 0 \\
 1 \\
-1 \\
-1 \\
-1 \\
\end{array}
\right)+{s\over9}
\left(
\begin{array}{c}
 1 \\
 8 \\
 -1 \\
 0 \\
-9 \\
\end{array}
\right)
\eea
\subsubsection{Anomaly $\mathcal A_3 (U(1)_A-[U(1)_Y]^2)$}
This anomaly condition puts constraints on the hypercharges of the exotic fields. 
The anomaly condition is give by 
\beq
q+8\,u+2\,d+3\,l-6\,e+h_1+h_2-6\,s \Sigma_iy_i^2 = 0 \label{A3}
\eeq
By taking the combination of Eqs.  (\ref{A3}) + (\ref{A2}) - 8 (\ref{yu}) + 2 (\ref{yd}) - 6 (\ref{yl}) + 6 (\ref{ys}), 
we get
\beq
\Sigma_i y_i^2=1
\eeq
which has several solutions. In the present work, we choose  $y_i = \{-{1\over3},{2\over3},{2\over3}\}$
\subsubsection{Anomalies $\mathcal A_4 (U(1)_Y-[U(1)_A]^2)$ \text{and}  $ A_5 [U(1)_A]^3$}
The final two anomalies do not have simple algebraic solutions. These are  given as 
$\mathcal A_4:$
\beq
3q^2-6u^2+3d^2-3l^2+3e^2-h_1^2 + ~h_2^2 + 3 ~\Sigma_i y_i(z_i^2-{\bar z_i}^2)=0 \label{A4}
\eeq
$\mathcal A_5:$
\beq
18q^3+9u^3+9d^3+6l^3+3e^3+2h_1^3+2h_2^3+s^3 + 3 ~\Sigma_i(z_i^3+{\bar z_i}^3)=0 \label{A5}
\eeq

We looked for integer solutions for the $U(1)_A$ charges. We could not find any as long as the charges are restricted
to lie below 10.  We then resorted to rational charges. There are several solutions which have been found. In Table
\ref{sam}, we present five sample solutions which satisfy the anomaly conditions as well as the superpotential requirements. 
In addition to this set of charges, one can also find sets where all the $z_i$ and $\bar{z}_i$ are equal.  It should also be noted
that each of the set of the charges has a completely different phenomenology. This is because the charges decide the $U(1)_A$
one loop beta function, $b_4$, which could vary drastically. This in turn modifies the values of $\lambda$ and $\kappa_i$ allowed
and their respective ranges. 

\begin {table}%
\caption{}
\begin {tabular}{|c| c| c|c | c|c|c| c|c| c| c|c | c| c |c |}
\toprule %
q& u& d &l &e &$h_1$&$h_2$ & s& $z_1$ &$z_2$&$z_3$& $\bar{z_1}$ &$\bar{z_2}$&$\bar{z_3}$ \\ \hline
${1\over 6}$& ${1\over 3}$& ${7\over 3}$&${1\over 2}$&${2}$&$-{5\over2}$&$-{1\over 2}$&${3}$&-3&${-1}$&${-1}$&${0}$&-2&-2 \\ [1ex]\hline
$-{1\over 18}$& $-{5\over 18}$& ${11\over 9}$&$-{3\over 2}$&$-{5\over6}$&${7\over6}$&${1\over 3}$&$-{3\over2}$&${1\over3}$&${1\over3}$&${1\over3}$&$-{7\over6}$&$-{7\over6}$&$-{7\over6}$ \\ [1ex]\hline
$-{1\over 27}$& ${10\over 27}$& $-{8\over 27}$&$-{1\over 3}$&${0}$&${1\over3}$&$-{1\over 3}$&${2\over3}$&$-{14 \over 27}$&$-{14\over 27}$&$-{14\over 27}$&${4\over 27}$&${4\over 27}$&${4\over 27}$ \\ [1ex]\hline
${1\over 27}$& ${5\over 27}$& $-{22\over 27}$&${2\over 9}$&${5\over9}$&$-{7\over9}$&$-{2\over 9}$&${1}$&-${2\over9}$&-${2\over9}$&-${2\over9}$&-${7\over9}$&-${7\over9}$&-${7\over9}$\\ [1ex]\hline
\end {tabular}
\label{sam}
\end {table}

\section{One loop corrections to the CP even Higgs mass matrix}
In the following we present the one loop corrections to the CP even Higgs mass matrix. There 
are two main contributions, one from the stop-top sector and the second one from from the vector
like exotic quarks.To derive the one loop
corrections, we use the well known  effective potential methods.   The one loop effective potential 
is given by \cite{Coleman:1973jx} 
\begin{eqnarray}
V^{1} & = &  \frac{3}{32 \pi^2} \left[ \sum_{j=1}^{2} m_{\widetilde{f}_{j}}^{4}\left( \ln \frac{m_{\widetilde{f}_{j}}^{2}}{Q^{2}}-\frac{3}{2}\right)
-2 \bar{m}_{f}^{4}\left( \ln \frac{\bar{m}_{f}^{2}}{Q^{2}}-\frac{3}{2}\right) \right].
\end{eqnarray}
where $m_{\widetilde{f}_{1,2}}^{2}$ are the  eigenvalues of the field dependent sfermion mass matrix. $\bar{m}_f$ is the corresponding
fermion mass. \\
The corrections to the CP even mass matrices can be written as  
\beq
\left( {\mathcal{M}}_{+}^{1}\right)_{ij}= \left. \frac{\partial ^{2}V^{1}}{\partial \phi _{i}\partial \phi_{j}} \right|_0 - \left. \delta_{ij} \frac{1}{v_i} \frac{\partial V^1}{\partial \phi_i} \right|_0
\eeq
By denoting 
\bea
\frac{\partial ^{2}m_{\widetilde{f}_{l}}^{2}}{\partial \phi _{i}\partial \phi_{j}}&=& {A}_{ij}^{\prime}\pm {A}_{ij}\nn
\frac{\partial m_{\widetilde{f}_{l}}^{2}}{\partial \phi_i}&=& {B}_{i}^{\prime}\pm {B}_{i}\nonumber
\eea 
mass matrix can be written as
\begin{eqnarray}
\left( {\mathcal{M}}_{+}^{1}\right)_{ij}&=& 2\,k\left[{\mathcal{F}}_{\tilde f} \,(A_{ij}^{\prime}-{\delta_{ij}\over H_j} B_j^{\prime})+{\mathcal{G}}_{\tilde f} \,(A_{ij}-{\delta_{ij}\over \phi_j} B_j)+{\mathcal{FF}}_{\tilde f}\,(B_i^{\prime}B_j^{\prime}+B_i B_j)+{\mathcal{GG}}_{\tilde f}\,(B_i^{\prime}B_j+B_i B_j^{\prime})\right.\nn
&&\left.-8\,{\mathcal{H}}_{ f}\, y_f^4 \,\langle\phi\rangle^2 \right ]
\label{genmat}
\end{eqnarray}
where
\begin{eqnarray}
{\mathcal{F}}_{\tilde f}&=& -(m_{\tilde f_2}^2+m_{\tilde f_1}^2)+(m_{\tilde f_2}^2 \log{m_{\tilde f_2}^2\over Q^2}+m_{\tilde f_1}^2\log{m_{\tilde f_1}^2\over Q^2})\nn
{\mathcal{G}}_{\tilde f}&=& (m_{\tilde f_2}^2-m_{\tilde f_1}^2)+(m_{\tilde f_2}^2 \log{m_{\tilde f_2}^2\over Q^2}-m_{\tilde f_1}^2\log{m_{\tilde f_1}^2\over Q^2})\nn
{\mathcal{FF}}_{\tilde f}&=&\log{m_{\tilde f_1}^2m_{\tilde f_2}^2\over Q^4}\nn
{\mathcal{GG}}_{\tilde f}&=&\log{m_{\tilde f_2}^2\over m_{\tilde f_1}^2 }\nn
{\mathcal{H}}_{\tilde f}&=&\log{m_{ f}^2\over Q^2}\nonumber
\end{eqnarray}
\hspace{4 cm}and {\large $k={3\over 32 \pi^2}$}\\
To include corrections to the Higgs mass matrix from the stop-top loop and all the three exotic quarks, we need to calculate \ref{genmat} in each case separately and add them. We have presented below corrections from the stop-top loop and one exotic quark.
\subsection{Top-Stop correction}
Dominant one loop correction to the Higgs mass matrix comes from the top and stop loop. The stop mass squared matrix is given as
\bea
\mathcal{M}_{\widetilde t}^2 = \left(\begin{array}{cc} M_{\widetilde{Q}}^{2}+y_t^{2} |H_2|^2  &X_t\\
(X_t)^{\dagger} & M_{\widetilde{U}}^{2}+y_t^{2} |H_2|^2\end{array}\right),
\eea
where $X_t=({A_t H_2}-\mu{\rm_ {eff}} H_1 y_t )$ and $m_t=y_t H_2$ 
%
\begin{eqnarray}
A_{11}&=&\mu_{eff}^2 y_t^2 \left[{2\over \est} -{ 8 X_{t}^2\over {\est}^3}\right]\nn
A_{12}&=&-\,\mu_{eff}\,y_t\,A_t\left[{2\over \est} -{ 8 X_{t}^2\over {\est}^3}\right]\nn
A_{13}&=&\left[\frac{-2 A_t H_2 \lambda y_t}{\est}-\frac{8 X_t^2 \mu_{eff} \lambda H_1 y_t^2}{{\est}}^3\right]\nn
A_{22}&=& A_t^2  \left[{2\over \est} -{ 8 X_{t}^2\over {\est}^3}\right]\nn
A_{23}&=& -\mu_{eff}y_t A_t \left[{2\over \est} -{ 8 X_{t}^2\over {\est}^3}\right]\nn
A_{33}&=&{\lambda^2y_t^2 H_1^2 }\left[{2\over \est} -{ 8 X_{t}^2\over {\est}^3}\right]\nn
A_{i2}^{\prime}&=&\delta_{i2} 2 y_t^2 \nn
B_{1}&=&{-2\,X_t\,\mu_{eff}\,y_t\over \est}\nn
B_{2}&=& {2\,X_t\,A_t\over \est}\nn
B_{3}&=&{-2\,X_t\,\lambda H_1\,y_t\over \est}\nn
B_{i}^{\prime}&=&\delta_{i2}=2 y_t^2 H_2\nonumber
\end{eqnarray}
\subsection{Correction due to Exotic quarks}
 The one loop correction due to the exotic quarks changes Higgs mass by few percent. The  exotic quark mass matrix given by 
\bea
\mathcal{M}_{\widetilde D_i}^2 = \left(\begin{array}{cc} M_{\tilde{D_i}}^{2}+k_i^{2} |S|^2  &X_{d_i}\\
(X_{d_i})^{\dagger} & M_{\widetilde{\bar D_i}}^{2}+k_i^{2} |S|^2\end{array}\right),
\eea
where $X_{d_i}=({A_{k_i} S}-\lambda\, k_i H_1 H_2 )$ and\,
   $ m_{D_i}=k_i S$
\begin{eqnarray}
A_{11}&=& (\lambda k_i H_2)^2 \left[{2\over \esd} -{ 8 X_{d_i}^2\over {\esd}^3}\right]\nn
A_{22}&=& (\lambda k_i H_1)^2 \left[{2\over \esd} -{ 8 X_{d_i}^2\over {\esd}^3}\right]\nn
A_{33} &=& A_{k_i}^2 \left[{2\over \esd} -{ 8 X_{d_i}^2\over {\esd}^3}\right]\nn
A_{12}&=&\left[ {\lambda^2 k_i^2 H_1 H_2 - 2 \lambda k_i A_{k_i} S\over \esd} - 
 {2 X_{d_i}^2 \lambda^2 k_i^2\,H_1 H_2\over {\esd}^3}\right]\nn
A_{13}&=&\lambda k_i H_2 A_{k_i} \left[{2\over \esd} -{ 8 X_{d_i}^2\over {\esd}^3}\right]\nn
A_{23}&=&\lambda k_i H_1 A_{k_i} \left[{2\over \esd} -{ 8 X_{d_i}^2\over {\esd}^3}\right]\nn
A_{i3}^{\prime}&=&\delta_{i3}\,2 k_i^2\nn
B_1&=&-2\,\lambda k_i H_2 {X_{d_i}\over\esd}\nn
B_2&=&-2\,\lambda k_i H_1 {X_{d_i}\over\esd}\nn
B_3&=& 2 A_{k_i} {X_{d_i}\over\esd}\nn
B_i^{\prime}&=&\delta_{i3}2 k_i^2 S\nonumber
\end{eqnarray}   

\section{RG Equations}
In the last section of the appendix we present the renormalisation equations for the various superpotential
and gauge parameters as well as soft terms. To derive the formulae we use the standard formulae available
in the literature\cite{Falck:1985aa,Martin:1993zk}.  The notation we use is $t=Log({\mu\over M_{susy}})$. 

\label{rge}
\bea
\frac{d g_i}{dt}&=&\frac{1}{16 \pi^2}{\beta_i}^{(1)}+\frac{1}{(16 \pi^2)^2}{\beta_i}^{(2)}\\
\frac{d y_i}{dt}&=&\frac{y_i}{16 \pi^2}{\gamma_i}^{(1)}+\frac{y_i}{(16 \pi^2)^2}{\gamma_i}^{(2)}
\eea
 \beq
\beta_a^{(1)} = b_{a} g_a^3,
\eeq
where $~b_a =~ \{ 17, 1, 0\}$ and\,
$b4=18q^2+6l^2+9(u^2+d^2)+3e^2+s^2+2(h_1^2+h_2^2)+3(z_1^2+z_2^2+z_3^2+(s+z_1)^2+(s+z_2)^2+(s+z_3)^2)$,
\bea
\cr{\beta_1}^{(2)}&=&4g_1^3\left(\frac{287}{36}g_1^2+\frac{9}{4}g_2^2+\frac{46}{3}g_3^2+(q^2/2+d^2+4u^2+3l^2/2+(h_1^2+h_2^2)/2+3e^2\right.\nn
&&\left.+{1\over3}(z_1^2+(s+z_1)^2+4(z_2^2+z_3^2+(s+z_2)^2+(s+z_3)^2))-{1\over4}(\frac{26}{3}y_t^2+\frac{14}{3}y_b^2+6y_{\tau}^2\right.\nn
&&\left.+2\lambda^2+\frac{4}{3}k_1^2+\frac{16}{3}(k_2^2+k_3^2))\right)
\cr{\beta_2}^{(2)}&=& 4g_2^5+g_2^3\big(3g_1^2+4g_2^2+24g_3^2+g_4^2(18q^2+6l^2+4(h_1^2+h_2^2))-6(y_t^2+y_b^2)-2(y_{\tau}^2+\lambda^2))\nn
\cr{\beta_3}^{(2)}&=&-54g_3^5+4g_3^3\left(\frac{47}{12}g_1^2+\frac{9}{4}g_2^2+21g_3^2+g_4^2(3q^2+{3\over2}(u^2+d^2)+{1\over2}(z_1^2+z_2^2\right.\nn
&&\left.+z_3^2+(s+z_1)^2+(s+z_2)^2+(s+z_3)^2)-4(y_t^2+y_b^2)-{4\over3}\lambda^2-3(k_1^2+k_2^2+k_3^2))\right)\nn
\cr{\beta_4}^{(2)}&=&4g_4^3\left(g_1^2({q^2\over2}+4u^2+d^2+{3l^2\over2}+3e^2+{1\over2}(h_1^2+h_2^2)+{3\over9}(z_1^2+(s+z_1)^2+4(z_2^2+z_3^2+(s+z_2)^2\right.\nn
&&\left.+(s+z_3)^2))+g_2^2({27\over2}q^2+{9\over2}l^2+{3\over2}(h_1^2+h_2^2))+g_3^2(24q^2+12(u^2+d^2)+4(z_1^2+z_2^2+z_3^2+(s+z_1)^2\right.\nn
&&\left.+(s+z_2)^2+(s+z_3)^2))+g_4^2(18q^4+9(u^4+d^4)+6l^4+3e^4+2(h_1^4+h_2^4)+s^4+3(z_1^4+z_2^4+z_3^4\right.\nn
&&\left.+(s+z_1)^4+(s+z_2)^4+(s+z_3)^4))-{1\over4}(12y_t^2(q^2+u^2+h_2^2)+12y_b^2(q^2+d^2+h_1^2)+4y_{\tau}^2(l^2+e^2+h_1^2)\right.\nn
&&\left.+4\lambda^2(s^2+h_1^2+h_2^2)+6k_1^2(s^2+z_1^2+(s+z_1)^2)+6k_2^2(s^2+z_2^2+(s+z_2)^2)+6k_3^2(s^2+z_3^2+(s+z_3)^2)))\right)\nonumber
\eea
\bea
\cr{\gamma_t}^{(1)}&=&\left[\lambda ^2+6 y_t^2 + y_b^2-{16\over 3} g_3^2 - 3 g_2^2 - {13\over 9} g_1^2 -2 g_4^2(q^2+u^2+h_2^2)\right] \nn
\cr{\gamma_b}^{(1)}&=&\left[\lambda ^2+6 y_b^2 + y_t^2+{y_\tau}^2-{16\over 3} g_3^2 - 3 g_2^2 - {7\over 9} g_1^2 -2 g_4^2(q^2+d^2+h_1^2)\right] \nn
\cr{\gamma_{\tau}}^{(1)}&=&\left[\lambda ^2+3 y_b^2 +4{y_\tau}^2 - 3 g_2^2 -3 g_1^2 -2 g_4^2(l^2+e^2+h_1^2)\right] \nn
\cr{\gamma_{\lambda}}^{(1)}&=&\left[4 \lambda ^2+3 (k_1^2+k_2^2+k_3^2)+3(y_t^2+y_b^2)+{y_\tau}^2-g_1^2 -2 g_4^2(s^2+h_2^2+h_1^2)\right] \nn
\cr{\gamma_{k_1}}^{(1)}&=&\left[2 \lambda ^2+5 k_1^2-{16\over 3} g_3^2- {4\over 9} g_1^2 -2 g4^2(s^2+z_1^2+(s+z_1)^2)\right] \nn
\cr{\gamma_{k_2}}^{(1)}&=&\left[2 \lambda ^2+5 k_2^2-{16\over 3} g_3^2- {8\over 9} g_1^2 -2 g4^2(s^2+z_2^2+(s+z_2)^2)\right] \nn
\cr{\gamma_{k_3}}^{(1)}&=&\left[2 \lambda ^2+5 k_3^2-{16\over 3} g_3^2- {8\over 9} g_1^2 -2 g4^2(s^2+z_3^2+(s+z_3)^2)\right]  \nn
\cr{\gamma_t}^{(2)}&=&\left[-22 y_t^4-5 y_b^4-y_t^2(3 \lambda^2+5 y_b^2)-y_b^2 y_{\tau}^2-3 \lambda^4-4 y_b^2 \lambda^2-\lambda^2 y_{\tau}^2 \right. \nn
&&\left.-3 \lambda^2(k_1^2+k_2^2+k_3^2)+y_t^2(2 g_1^2+6g_2^2+16 g_3^2+g_4^2(8 q^2+4 u^2))+y_b^2({2\over3}g_1^2+2g_4^2(d^2\right. \nn
&&\left.+h_1^2-q^2))+2 \lambda^2 g4^2(h_1^2+s^2-h_2^2)+\frac{3679}{162}g_1^4+\frac{15}{2}g_2^4+\frac{416}{9}g_3^4+g_4^4(2s_4(q^2+u^2+h_2^2)\right.\nn
&&\left.+4(q^4+u^4+h_2^4))+\frac{5}{3}g_1^2g_2^2+\frac{136}{27}g_1^2g_3^2+8({h_2^2\over4}+{q^2\over36}+{4u^2\over9})g_1^2g_4^2+8g_2^2g_3^2+6g_2^2g_4^2\right. \nn
&&\left.(q^2+h_2^2)+\frac{32}{3}(q^2+u^2)g_3^2g_4^2\right]\nn
\cr{\gamma_b}^{(2)}&=&\left[-22 y_b^4-5 y_t^4-4y_t^2\lambda^2-y_b^2(3\lambda^2+5y_t^2+y_{\tau}^2-3\lambda^4-3y_{\tau}^4-3\lambda^2(k_1^2+k_2^2+k_3^2)\right.\nn
&&\left.+y_b^2({4\over3}g_1^2+{9\over2}g_2^2+16g_3^2+g_4^2(6 q^2+6d^2+2h_1^2))+2y_t^2({4\over3}g_1^2+2g_4^2(u^2+h_2^2-q^2))+2\lambda^2g_4^2(s^2\right.\nn
&&\left.+h_2^2-h_1^2)+y_{\tau}^2(2g_1^2+2g_4^2(l^2+e^2-h_1^2))+\frac{1939}{162}g_1^4+\frac{15}{2}g_2^4+\frac{416}{9}g_3^4+g_4^4(2s_4(q^2+d^2+h_1^2)\right.\nn
&&\left.+4(q^4+d^4+h_1^4))+\frac{5}{3}g_1^2g_2^2+\frac{40}{27}g_1^2g_3^2+8({h_1^2\over4}+{q^2\over36}+{4d^2\over9})g_1^2g_4^2+8g_2^2g_3^2+6g_2^2g_4^2(q^2+h_1^2)\right.\nn
&&\left.+\frac{32}{3}(q^2+d^2)g_3^2g_4^2\right]\nn
\cr{\gamma_{\tau}}^{(2)}&=&\left[-9y_b^4-3\lambda^4-10y_{\tau}^4-3y_t^2y_b^2-3\lambda^2y_t^2-3_{\tau}^2(\lambda^2+3y_b^2)-3\lambda^2(k_1^2+k_2^2+k_3^2)+y_b^2(-{2\over3}g_1^2+16g_3^2\right.\nn
&&\left.+6g_4^2(q^2+d^2-h_1^2))+2\lambda^2g_4^2(s^2+h_2^2-h_1^2)+y_{\tau}^2(2g_1^2+6g_2^2+4g_4^2(l^2+h_1^2))+\frac{99}{2}g_1^4\right.\nn
&&\left.+\frac{15}{2}g_2^4+g_4^4(2s_4(e^2+l^2+h_1^2)+4(l^4+h_1^4+e^4))+3g_1^2g_2^2+g_1^2g_4^2(2h_1^2+2l^2+8e^2)+6g_2^2g_4^2(h_1^2+l^2)\right]\nn
\cr{\gamma_{\lambda}}^{(2)}&=&\left[-9y_t^4-9y_b^4-10\lambda^4-3y_{\tau}^4-6y_t^2y_b^2-\lambda^2(9y_b^2+9y_t^2+3y_{\tau}^2+6(k_1^2+k_2^2+k_3^2))-6(k_1^4+k_2^4+k_3^4)\right.\nn
&&\left.+y_t^2({3\over2}g_1^2+16g_3^2+6g_4^2(u^2+q^2-h_2^2))+\lambda^2(2g_1^2+6g_2^2+4g_4^2(h_1^2+h_2^2))+k_1^2({4\over3}g_1^2\right.\nn
&&\left.+16g_3^2+6g_4^2(z_1^2+(s+z_1)^2-s^2))+k_2^2({16\over3}g_1^2+16g_3^2+6g_4^2(z_2^2+(s+z_2)^2-s^2))+k_3^2({16\over3}g_1^2\right.\nn
&&\left.+16g_3^2+6g_4^2(z_3^2+(s+z_3)^2-s^2))+y_b^2(-{2\over3}g_1^2+3g_2^2+16g_3^2+6g_4^2(q^2+d^2-h_1^2))+2y_{\tau}^2(g_1^2+g_4^2\right.\nn
&&\left.(l^2+e^2-h_1^2))+\frac{34}{3}g_1^4+\frac{15}{2}g_2^4+g_4^4(2s_4(h_1^2+s^2+h_2^2)+4(h_1^4+s^4+h_2^4))+3g_1^2g_2^2\right.\nn
&&\left.+2g_1^2g_4^2(h_1^2+h_2^2)+6g_2^2g_4^2(h_1^2+h_2^2)\right]\nn
\cr{\gamma_{k_1}}^{(2)}&=&\left[-6k_1^2\lambda^2-6k_1^4-4\lambda^4-\lambda^2(2y_{\tau}^2+6y_b^2+6y_t^2)-6k_1^2(k_1^2+k_2^2+k_3^2)+k_1^2({4\over3}g_1^2+16g_3^2+2g_4^2(z_1^2\right.\nn
&&\left.+(s+z_1)^2-s^2)+\lambda^2(2g_1^2+6g_2^2+2g_4^2(h_1^2+h_2^2-s^2))+\frac{542}{81}g_1^4+\frac{416}{9}g_3^4+g_4^4(2s_4(z_1^2\right.\nn
&&\left.+(s+z_1)^2)+4(z_1^4+(s+z_1)^4))+\frac{64}{27}g_1^2g_3^2+\frac{8}{9}(z_1^2+(s+z_1)^2)g_1^2g_4^2+\frac{32}{3}(z_1^2+(s+z_1)^2)g_4^2g_3^2\right]\nn
\cr{\gamma_{k_2}}^{(2)}&=&\left[-6k_2^2\lambda^2-6k_2^4-4\lambda^4-\lambda^2(2y_{\tau}^2+6y_b^2+6y_t^2)-6k_2^2(k_1^2+k_2^2+k_3^2)+k_2^2({16\over3}g_1^2+16g_3^2+2g_4^2(z_2^2\right.\nn
&&\left.+(s+z_2)^2-s^2)+\lambda^2(2g_1^2+6g_2^2+2g_4^2(h_1^2+h_2^2-s^2))+\frac{2168}{81}g_1^4+\frac{416}{9}g_3^4+g_4^4(2s_4(z_2^2\right.\nn
&&\left.+(s+z_2)^2)+4(z_2^4+(s+z_2)^4))+\frac{256}{27}g_1^2g_3^2+\frac{32}{9}(z_2^2+(s+z_2)^2)g_1^2g_4^2+\frac{32}{3}(z_2^2+(s+z_2)^2)g_4^2g_3^2\right]\nn
\cr{\gamma_{k_3}}^{(2)}&=&\left[-6k_3^2\lambda^2-6k_3^4-4\lambda^4-\lambda^2(2y_{\tau}^2+6y_b^2+6y_t^2)-6k_3^2(k_1^2+k_2^2+k_3^2)+k_3^2({16\over3}g_1^2+16g_3^2+2g_4^2(z_3^2\right.\nn
&&\left.+(s+z_3)^2-s^2)+\lambda^2(2g_1^2+6g_2^2+2g_4^2(h_1^2+h_2^2-s^2))+\frac{2168}{81}g_1^4+\frac{416}{9}g_3^4+g_4^4(2s_4(z_3^2\right.\nn
&&\left.+(s+z_3)^2)+4(z_3^4+(s+z_3)^4))+\frac{256}{27}g_1^2g_3^2+\frac{32}{9}(z_3^2+(s+z_3)^2)g_1^2g_4^2+\frac{32}{3}(z_3^2+(s+z_3)^2)g_4^2g_3^2\right]\nonumber
\eea
where $s_4=18q^2+9(u^2+d^2)+6l^2+2(h_1^2+h_2^2)+3(z_1^2+z_2^2+z_3^2+(s+z_1)^2+(s+z_2)^2+(s+z_3)^2)+e^2+s^2$,

\bea
\cr\frac{d m_{q_3}^2}{dt} &=& \frac{1}{16 \pi^2}\left[2y_t^2(m_{q_3}^2+m_{u_3}^2+m_2^2)+2A_t^2+2y_b^2(m_{q_3}^2+m_{d_3}^2+m_1^2)+2A_b^2\right.\nn
&&\left.-{32\over 3}g_3^2M_3^2-6g_2^2M_2^2-{2\over 9}g_1^2M_1^2-8q^2g_4^2M_4^2+{1\over 3}g_1^2\xi+2qg_4^2\xi^\prime\right]\nn
\cr\frac{d m_{u_3}^2}{dt} &=& \frac{1}{16 \pi^2}\left[4y_t^2(m_{q_3}^2+m_{u_3}^2+m_2^2)+4A_t^2-{32\over 3}g_3^2M_3^2-8{4\over 9}g_1^2M_1^2-8u^2g_4^2M_4^2-{4\over 3}g_1^2\xi+2ug_4^2\xi^\prime\right]
\cr\frac{d m_{d_3}^2}{dt} &=& \frac{1}{16 \pi^2}\left[4y_b^2(m_{q_3}^2+m_{d_3}^2+m_1^2)+4A_b^2-{32\over 3}g_3^2M_3^2-8{1\over 9}g_1^2M_1^2-8d^2g_4^2M_4^2+{2\over 3}g_1^2\xi+2dg_4^2\xi^\prime\right]
\cr\frac{d m_{l_3}^2}{dt} &=& \frac{1}{16 \pi^2}\left[2y_\tau^2(m_{l_3}^2+m_{\tau}^2+m_1^2)+2A_\tau^2-6g_2^2M_2^2-8{1\over 4}g_1^2M_1^2-8l^2g_4^2M_4^2-g_1^2\xi+2lg_4^2\xi^\prime\right]
\cr\frac{d m_\tau^2}{dt} &=& \frac{1}{16 \pi^2}\left[4y_\tau^2(m_{l_3}^2+m_{\tau}^2+m_1^2)+4A_\tau^2-8g_1^2M_1^2-8e^2g_4^2M_4^2+2g_1^2\xi+2eg_4^2\xi^\prime\right]\nn
\cr\frac{d m_1^2}{dt} &=& \frac{1}{16 \pi^2}\left[6y_b^2(m_{d_3}^2+m_{q_3}^2+m_1^2)+6A_b^2+2y_\tau^2(m_\tau^2+m_{l_3}^2+m_1^2)+2A\tau \right.\nn
&&\left.+2\lambda^2(m_1^2+m_2^2+m_s^2)+2A\lambda-6g_2^2M_2^2-2g_1^2M_1^2-8h_1^2g_4^2M_4^2-g_1^2\xi+2h_1g_4^2\xi^\prime\right]\nn
\cr\frac{d m_2^2}{dt} &=& \frac{1}{16 \pi^2}\left[6y_t^2(m_{u_3}^2+m_{q_3}^2+m_2^2)+6A_t^2+2\lambda^2(m_1^2+m_2^2+m_s^2)\right.\nn
&&\left.+2A\lambda-6g_2^2M_2^2-2g_1^2M_1^2-8h_1^2g_4^2M_4^2+g_1^2\xi+2h_1g_4^2\xi^\prime\right]\nn
\cr\frac{d m_s^2}{dt} &=& \frac{1}{16 \pi^2}\left[6 k_1^2(m_s^2+m_{D_1}^2+m_{{\bar D}_1}^2)+6A_{k_1}^2+6 k_2^2(m_s^2+m_{D_2}^2+m_{{\bar D}_2}^2)\right.\nn
&&\left.+6A_{k_2}^2+6 k_3^2(m_s^2+m_{D_3}^2+m_{{\bar D}_3}^2)+6A_{k_3}^2+4\lambda^2(m_1^2+m_2^2+m_s^2)+4A\lambda-8s^2g_4^2M_4^2+2sg_4^2\xi^\prime\right]\nn
\cr\frac{d m_{D_1}^2}{dt} &=& \frac{1}{16 \pi^2}\left[2 k_1^2(m_s^2+m_{D_1}^2+m_{{\bar D}_1}^2)+2A_{k_1}^2-{32\over 3}g_3^2M_3^2-{8\over 9}g_1^2 M_1^2-8z_1^2g_4^2M_4^2-{2\over 3}g_1^2\xi+2z_1g_4^2\xi^\prime\right]\nn
\cr\frac{d m_{D_2}^2}{dt} &=& \frac{1}{16 \pi^2}\left[2 k_2^2(m_s^2+m_{D_2}^2+m_{{\bar D}_2}^2)+2A_{k_2}^2-{32\over 3}g_3^2M_3^2-{32\over 9}g_1^2 M_1^2-8z_2^2g_4^2M_4^2+{4\over 3}g_1^2\xi+2z_2g_4^2\xi^\prime\right]\nn
\cr\frac{d m_{D_3}^2}{dt} &=& \frac{1}{16 \pi^2}\left[2 k_3^2(m_s^2+m_{D_3}^2+m_{{\bar D}_3}^2)+2A_{k_3}^2-{32\over 3}g_3^2M_3^2-{32\over 9}g_1^2 M_1^2-8z_3^2g_4^2M_4^2+{4\over 3}g_1^2\xi+2z_3g_4^2\xi^\prime\right]
\cr\frac{d m_{{\bar D}_1}^2}{dt} &=& \frac{1}{16 \pi^2}\left[2 k_1^2(m_s^2+m_{D_1}^2+m_{{\bar D}_1}^2)+2A_{k_1}^2-{32\over 3}g_3^2M_3^2-{8\over 9}g_1^2 M_1^2-8(s+z_1)^2g_4^2M_4^2+{2\over 3}g_1^2\xi\right.\nn
&&\left.+2(s+z_1g_4^2\xi^\prime\right]\nn
\cr\frac{d m_{{\bar D}_2}^2}{dt} &=& \frac{1}{16 \pi^2}\left[2 k_2^2(m_s^2+m_{D_2}^2+m_{{\bar D}_2}^2)+2A_{k_2}^2-{32\over 3}g_3^2M_3^2-{32\over 9}g_1^2 M_1^2-8(s+z_2)^2g_4^2M_4^2-{4\over 3}g_1^2\xi\right.\nn
&&\left.+2(s+z_2)g_4^2\xi^\prime\right]\nn
\cr\frac{d m_{{\bar D}_3}^2}{dt} &=& \frac{1}{16 \pi^2}\left[2 k_3^2(m_s^2+m_{D_3}^2+m_{{\bar D}_3}^2)+2A_{k_3}^2-{32\over 3}g_3^2M_3^2-{32\over 9}g_1^2 M_1^2-8(s+z_3)^2g_4^2M_4^2-{4\over 3}g_1^2\xi\right.\nn
&&\left.+2(s+z_3)g_4^2\xi^\prime\right]\nonumber
\eea
\bea
\frac{d A_t}{dt} &=& \frac{A_t}{16 \pi^2}\left[18y_t^2+y_b^2+\lambda^2-{16\over3}g_3^2-3g_2^2-{13\over9}g_1^2-2(q^2+u^2+h2^2)g_4^2\right]\nn&&+\frac{y_t}{16 \pi^2}\left[2y_bA_b+A_{\lambda}\lambda+{32\over3}g_3^2M_3+6g_2^2M_2+{26\over9}g_1^2M_1+4(q^2+u^2+h_2^2)g_4^2M_4\right]\nn
\cr\frac{d A_b}{dt} &=& \frac{A_b}{16 \pi^2}\left[18y_b^2+y_t^2+y_\tau^2+\lambda^2-{16\over3}g_3^2-3g_2^2-{7\over9}g_1^2-2(q^2+d^2+h_1^2)g_4^2\right]\nn&&+\frac{y_b}{16 \pi^2}\left[2y_tA_t+2A_\tau y_\tau+2A_{\lambda}\lambda+{32\over3}g_3^2M_3+6g_2^2M_2+{14\over9}g_1^2M_1+4(q^2+d^2+h_1^2)g_4^2M_4\right]\nn
\cr\frac{d A_\tau}{dt} &=& \frac{A_\tau}{16 \pi^2}\left[12y_\tau^2+3y_b^2+\lambda^2-3g_2^2-3g_1^2-2(l^2+e^2+h1^2)g_4^2\right]\nn&&+\frac{y_tau}{16 \pi^2}\left[6y_bA_b+2A_{\lambda}\lambda+6g_2^2M_2+6g_1^2M_1+4(l^2+e^2+h1^2)g_4^2M_4\right]\nn
\cr\frac{d A_\lambda}{dt} &=& \frac{A_\lambda}{16 \pi^2}\left[3y_b^2+3y_t^2+y_\tau^2+12\lambda^2+3(k_1^2+k_2^2+k_3^2)-3g_2^2-g_1^2-2(s^2+h_2^2+h_1^2)g_4^2\right]\nn&&+\frac{\lambda}{16 \pi^2}\left[6y_tA_t+6y_bA_b+2A_\tau y_\tau+6(A_{k_1}k_1+A_{k_2}k_2+A_{k_3}k_3)\right.\nn
&&\left.+6g_2^2M_2+2g_1^2M_1+4(s^2+h_2^2+h1^2)g_4^2M_4\right]\nn
\cr\frac{d A_{k_1}}{dt} &=& \frac{A_{k_1}}{16 \pi^2}\left[3k_1^2+\lambda^2-{16\over3}g_3^2-{4\over9}g_1^2-2(s^2+z_1^2+(s+z_1)^2)g_4^2\right]\nn&&+\frac{k_1}{16 \pi^2}\left[4\lambda A_\lambda+{32\over3}g_3^2M_3+{8\over9}g_1^2M_1+4(s^2+z_1^2+(s+z_1)^2)g_4^2M_4\right]\nn
\cr\frac{d A_{k_2}}{dt} &=& \frac{A_{k_2}}{16 \pi^2}\left[3k_2^2+\lambda^2-{16\over3}g_3^2-{16\over9}g_1^2-2(s^2+z_2^2+(s+z_2)^2)g_4^2\right]\nn&&+\frac{k_2}{16 \pi^2}\left[4\lambda A_\lambda+{32\over3}g_3^2M_3+{32\over9}g_1^2M_1+4(s^2+z_2^2+(s+z_3)^2)g_4^2M_4\right]\nn
\cr\frac{d A_{k_3}}{dt} &=& \frac{A_{k_3}}{16 \pi^2}\left[3k_3^2+\lambda^2-{16\over3}g_3^2-{16\over9}g_1^2-2(s^2+z_3^2+(s+z_3)^2)g_4^2\right]\nn&&+\frac{k_3}{16 \pi^2}\left[4\lambda A_\lambda+{32\over3}g_3^2M_3+{32\over9}g_1^2M_1+4(s^2+z_2^2+(s+z_3)^2)g_4^2M_4\right]\nonumber
\eea

\bibliographystyle{ieeetr}
\bibliography{GMSB1.bib}
\end{document}